\documentclass{revtex4}
\usepackage{graphicx}
\usepackage{fancyhdr}
\usepackage{amsmath}
\usepackage{color}

\def\GeV{\rm{GeV}}
\def\TeV{\rm{TeV}}

\begin{document}

\title{Higgs phenomenology in the supersymmetric grand unified theory with the Hosotani mechanism\footnote{This talk is based on the work in Ref.\cite{SGGHU}}}

\author{Hiroyuki Taniguchi}
\affiliation{Department of Physics, University of Toyama, 3190 Gofuku, Toyama 930-8555, Japan}
\begin{abstract}
The supersymmetric grand unified theory with the Hosotani mechanism predicts the existence of adjoint chiral supermultiplets at the SUSY breaking scale. 
In particular, the $SU(2)$ triplet and the singlet chiral superfields affect the Higgs sector. 
We investigate the contributions from these adjoint chiral multiplets to the masses of the Higgs sector particles and their couplings to the standard model particles. 
We show that the predicted values of the Higgs sector parameters deviate from the standard model and the minimal supersymmetric standard model by $O(1)~\% -O(10)~\%$. 
\end{abstract}

\maketitle

\thispagestyle{fancy}


\section{Introduction}
Since the existence of a standard-model-like Higgs boson whose mass is around $126~{\rm GeV}$ was confirmed\cite{Aad:2012tfa}, the standard model (SM) is established as a low energy effective theory. 
The SM predictions are consistent with almost all observations. 
However, the SM has some problems such as the hierarchy problem, and the charge quantization is mysteries. 
These problems should be solved in the extensions of the SM. 

The grand unified theories (GUTs) unify the gauge groups in the SM and quantize the electric charge\cite{Georgi:1974sy}. 
The supersymmetery (SUSY) prevents the quadratic divergence from the Higgs boson mass and stabilizes the hierarchy between the electroweak scale and the cutoff scale. 
Therefore, SUSY-GUTs are well-motivated models of beyond the SM\cite{Witten:1981nf}. 
However, in the SUSY-GUTs, the GUT breaking scale is typically $O(10^{16})~\GeV$ as is inferred from the gauge coupling unification. 
Due to the decoupling theorem\cite{Appelquist:1974tg}, it is difficult to test the SUSY-GUTs at collider experiments. 
Tests of the SUSY-GUTs rely on checking the relations among the parameters of superparticles. 
There is another difficulty. 
In the SUSY-GUTs, the $SU(2)$ doublet Higgs fields necessarily accompany color triplet Higgs fields. 
The color triplet Higgs fields are as heavy as the GUT scale for proton longevity\cite{Nishino:2012ipa}, but the $SU(2)$ doublet Higgs fields should be around $O(10^2)~{\rm GeV}$ for the electroweak symmetry breaking. 
That is, the SUSY-GUTs also have a fine tuning problem that there is the mass splitting between the color triplet and the $SU(2)$ doublet Higgs fields which arise from common multiplet, so-called doublet-triplet (DT) splitting problem. 

In this situation, we consider the SUSY grand unified theory with the Hosotani mechanism\cite{Hosotani:1983xw}, so-called the SUSY grand Gauge-Higgs unification (SGGHU)\cite{Yamashita:2011an}. 
In this model, the GUT symmetry is broken by the Hosotani mechanism. 
The extra-dimensional component of the gauge field causes the symmetry breaking.
The SGGHU realizes naturally the DT splitting problem. 
Furthermore, this model predicts the existence of the adjoint chiral supermultiplets, the color octet, the $SU(2)$ triplet and the singlet, at the SUSY breaking scale. 

Since the $SU(2)$ triplet and the singlet are included in the Higgs sector, we can test the model by studying the Higgs sector at collider experiments. 
The predicted masses of the additional Higgs bosons and the Higgs couplings to the SM particles are also different from the other SUSY models. 
We will discuss the Higgs sector and show the verifiability of the GUTs by detecting the deviation from the SM and also from the minimal supersymmetric standard model (MSSM) at future collider experiments. 

\section{The Higgs sector of the SUSY grand gauge-Higgs unification}
\begin{figure}[h]
\includegraphics[width=80mm]{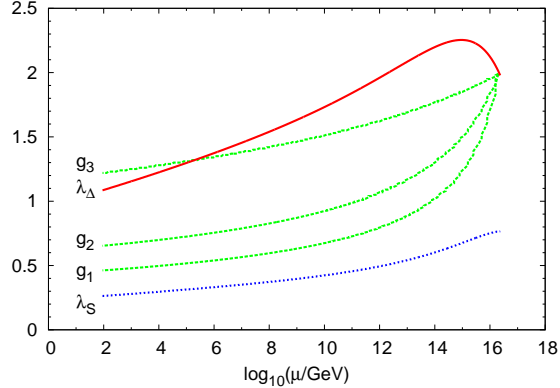}
\caption{The running of the gauge couplings, $\lambda_S^{}$(red line) and $\lambda_\Delta^{}$(blue line). At the GUT scale, $\lambda_\Delta^{}$ and $\lambda_S^{}$ are unified.
}
\label{fig:lambda}
\end{figure}

This model has the Higgs sector of the MSSM which is extended by the singlet $\hat{S}$ and the triplet $\hat{\Delta}$.
The superpotential is given as
\begin{align}
W = \mu \hat{H}_u \cdot \hat{H}_d + \mu_\Delta^{} {\rm Tr}{(\hat{\Delta}^2)} +\frac{\mu_S^{}}{2}\hat{S}^2 +\lambda_\Delta \hat{H}_u \cdot \hat{\Delta} \hat{H}_d +\lambda_S \hat{S} \hat{H}_u \cdot \hat{H}_d,
\end{align}
where $\hat{H_u}$, $\hat{H_d}$, $\hat{S}$ and $\hat{\Delta}$ have quantum numbers that are shown in Tab.~\ref{tb:fields}, and the hat represents superfield.
\begin{table}[h]
  \begin{tabular}{|c|c|c|c|c|}
    \hline
    &SU(3) &SU(2) &U(1) & \\ \hline \hline
    $\hat{H_u}$ &{\bf 1} &{\bf 2} &+1 &MSSM doublet \\ \hline
    $\hat{H_d}$ &{\bf 1} &{\bf 2} &-1 &MSSM doublet \\ \hline
    $\hat{S}$   &{\bf 1} &{\bf 1} &0 &Singlet \\ \hline
    $\hat{\Delta}$ &{\bf 1} &{\bf 3} &0 &Triplet \\ 
    \hline
  \end{tabular}
  \caption{Superfields of the Higgs sector.}
  \label{tb:fields}
\end{table}
Since $\hat{S}$ and $\hat{\Delta}$ are contained by the gauge field at the GUT scale, there is neither $\hat{S}\hat{S}\hat{S}$ nor $\hat{S}\hat{\Delta}\hat{\Delta}$ term.
Then the couplings $\lambda_S$ and $\lambda_\Delta$ are related to the gauge couplings at the GUT scale. 
This relationship is $\lambda_\Delta^{} =2\sqrt{5/3}\lambda_S^{}$.
In this talk, we use the low energy values of the couplings as
\begin{align}
\lambda_\Delta = 1.1, \hspace{1em} \lambda_S = 0.26,
\label{eq:couplings}
\end{align}
which are obtained by solving the renormalization group equations (Fig.~\ref{fig:lambda}).

The soft breaking term is written by
\begin{align}
V_{\rm soft} 
&=\tilde{m}_d^2 |H_d|^2 +\tilde{m}_u^2|H_u|^2 +\tilde{m}_\Delta^2|\Delta|^2 +\tilde{m}_S^2|S|^2 \nonumber \\
&+[B\mu H_u\cdot H_d +\eta S +B_\Delta^{}\mu_\Delta^{}{\rm Tr}(\Delta^2) +\frac{1}{2}B_S^{} \mu_S^{} S^2 \nonumber \\
&\hspace{1.5em} +\lambda_\Delta^{}A_\Delta^{} H_u\cdot \Delta H_d +\lambda_S A_S S H_u\cdot H_d +{\rm h.c.} ].
\end{align}
In this Higgs sector, there are 15 parameters.
The values of $\lambda_S$ and $\lambda_\Delta$ are obtained by solving the renormalization group equations, and 4 parameters are defined by the 4 tadpole conditions.
Therefore, this model has high predictability.
After the electroweak symmetry breaking, there are 3 physical CP-odd Higgs bosons, 4 CP-even Higgs bosons and 3 charged Higgs bosons.
Similarly to the next-to-MSSM (NMSSM\cite{Ellwanger:2009dp}), in this model, the SM-like Higgs boson mass is expected to be heavier by effects of trilinear coupling of Higgs, as compared to typical SUSY models\cite{Okada:1990vk}.
In the next section, we show that by measuring the masses and the coupling constants of the Higgs sector particles at the LHC and the ILC, we can distinguish the model.

\section{Phenomenology}
\begin{figure}[h]
\includegraphics[width=80mm]{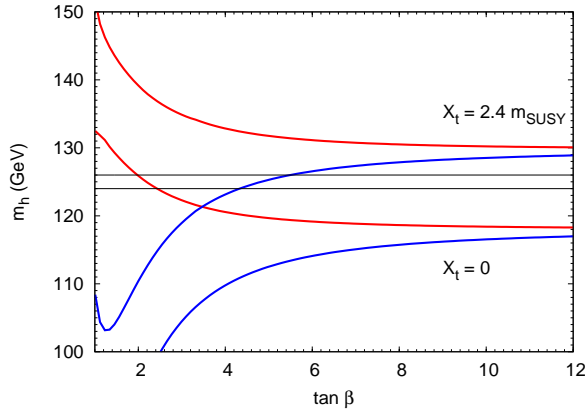}
\caption{The SM-like Higgs boson mass as a function of $\tan\beta$ for the large soft mass scenario $\tilde{m}_S^{}, \tilde{m}_\Delta^{} =2 ~{\rm TeV}$.  The red line is the result for the SGGHU case and the blue line is for the MSSM case. The upper line is for the maximum mixing case and the lower line is for the non-mixing case.}
\label{fig:mh_tb}
\end{figure}

First, we investigate the mass of the SM-like Higgs boson $h$.
The SM-like Higgs boson mass $m_h$ is written as\cite{Espinosa:1992hp}
\begin{align}
m_h^2 \sim 
m_Z^2 c_\beta^2 
+\frac{3 m_t^4}{2\pi^2 v^2}\left( \log{\frac{m_{\tilde{t}}^2}{m_t^2}} +\frac{X_t^2}{m_{\tilde{t}}^2}(1 -\frac{X_t^2}{12m_{\tilde{t}}^2}) \right)
+\frac{1}{2}\lambda_S^2 v^2 s_{2\beta}^2 +\frac{1}{8}\lambda_\Delta^2 v^2 s_{2\beta}^2,
\end{align}
where $s_\beta =\sin\beta$, $c_\beta =\cos\beta$, $\tan\beta =v_u/v_d$, $v^2 =(v_u^2 +v_d^2) \sim (246~{\rm GeV})^2$ and $X_t =A_t -\mu \cot\beta$ is stop mixing parameter.
Since $m_h$ is less than the Z boson mass at the tree level, the loop correction should be relatively large in order to reach $126~\GeV$ in the MSSM.
In the maximum mixing case $X_t =\pm \sqrt{6}m_{\rm SUSY}^{}$, where $m_{\rm SUSY}^{}$ is a scale of typical SUSY-parameters, the stop mass is $O(1)~\TeV$\cite{Hall:2011aa}.
However, in the non-mixing case $X_t =0$, the stop mass is $O(10)~\TeV$. 
It seems that the heavy stop which can explain the $126~\GeV$ Higgs boson is a kind of new fine tuning. 
On the contrary, in the SGGHU, $m_h$ becomes large as compared to the MSSM by the tree level F-term contributions. 
In other words, the small stop mass is allowed even in the non-mixing case.
Fig.~\ref{fig:mh_tb} shows $m_h$ as a function of $\tan\beta$ in the SGGHU and the MSSM for the large soft mass scenario $\tilde{m}_S^{}, \tilde{m}_\Delta^{} =2 ~{\rm TeV}$.
The red line is the result for the SGGHU case and the blue line is for the MSSM case.
The upper line and the lower line correspond to the maximum mixing case and the non-mixing case, respectively. 
$m_h$ can reach $126~{\rm GeV}$ at the low $\tan\beta$ in this model.

\begin{figure}[h]
\includegraphics[width=80mm]{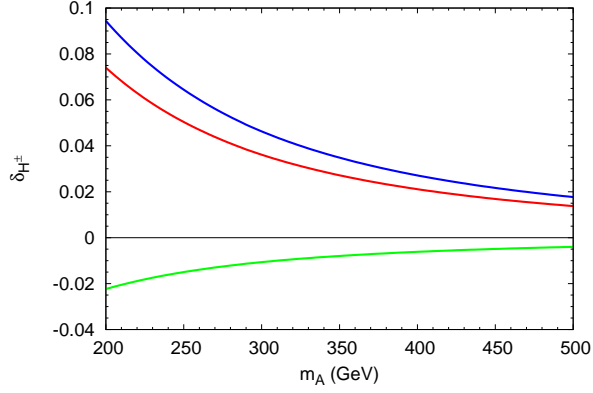}
\caption{The deviation in $m_{H^\pm}^{}$ from the MSSM as a function of the CP-odd Higgs boson mass $m_A$ in the large soft mass scenario. The green, blue and red line correspond to the NMSSM, the MSSM with triplet and the SGGHU case, respectively. }
\label{fig:dhpm_ma}
\end{figure}

Second, we discuss the deviation in the masses of heavy Higgs bosons from the MSSM predictions.
The charged Higgs boson mass $m_{H^\pm}^{}$ is given as 
\begin{align}
m_{H^\pm}^2 
&= m_{H^\pm}^2|_{\rm MSSM}^{} (1 +\delta_{H^\pm}^{})^2 \nonumber \\
&\sim m_A^2 +m_W^2 -\frac{1}{2}\lambda_S^2v^2 +\frac{1}{8}\lambda_\Delta^2 v^2,
\end{align}
where $\delta_{H^\pm}$ is the deviation in $m_{H^\pm}^{}$ from the MSSM and $m_A^{}$ is the CP-odd Higgs boson mass.
The singlet effect is opposite to the triplet effect by the group theory.
Due to Eq.~(\ref{eq:couplings}), $m_{H^\pm}^{}$ becomes large as compared to the MSSM. 
We emphasize that these $\lambda_S^{}$ and $\lambda_\Delta^{}$ couplings are determined by the renormalization group equations and large $m_{H^\pm}^{}$ is the prediction in this model. 
Since $m_{H^\pm}|_{\rm MSSM}^{}$ is the sum of $m_A^{}$ and $m_W^{}$, 
when the CP-odd Higgs boson and the charged Higgs boson are discovered, we can obtain $\delta_{H^\pm}$ by measuring the deviation between $m_A^{}$ and $m_{H^\pm}^{}$.
Fig.~\ref{fig:dhpm_ma} shows the deviation in $m_{H^\pm}^{}$ from the MSSM as a function of $m_A^{}$ in the large soft mass scenario.
The green, blue and red lines show the NMSSM, the MSSM with triplet and the SGGHU case, respectively.
The mass deviation is $O(1)~\%-O(10)~\%$.
On the other hand, the deviation in the heavy CP-even Higgs boson mass $m_H^{}$ from the MSSM prediction is less than $O(1)~\%$. 
Since the charged Higgs mass can be determined with an accuracy of a few percent at the LHC\cite{Cavalli:2002goa}, we can test the SGGHU.

\begin{figure}[h]
\includegraphics[width=80mm]{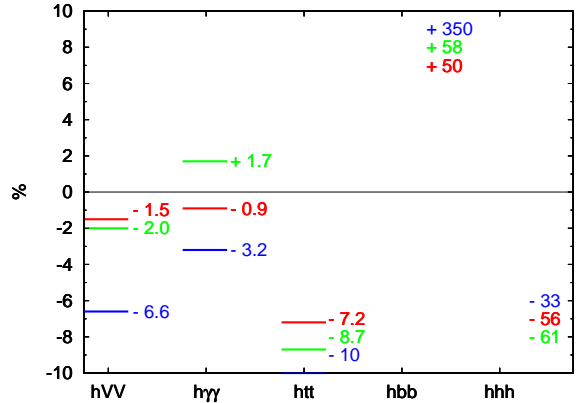}
\caption{The deviation in Higgs couplings from the SM at the leading order. 
Here the vertical axis is $g_{hXX}^{}/g_{hXX}^{}(SM) -1$ and $X =V(=W,Z),\gamma,t,b,h$. 
The blue, green and red lines show the results for the MSSM, the NMSSM and the SGGHU case, respectively. 
Although absolute values of the $hb\bar{b}$ and $hhh$ couplings are larger than $10~\%$, the Higgs-gauge boson couplings are less than $O(1)~\%$.
}
\label{fig:Higgscouplings}
\end{figure}

Then, we investigate the Higgs couplings to the SM particles. 
The SM-like Higgs boson was discovered, but no one knows the detail of this Higgs sector.
In order to construct a model of beyond the SM, it is important to study the deviation in the Higgs couplings from the SM values. 
From Fig.~\ref{fig:Higgscouplings}, the deviation in the Higgs couplings is $O(1)~\%$.
As Ref.\cite{Peskin:2012we} shown, the ILC can reach accuracies better than a few percent for the Higgs-gauge boson couplings.
Therefore, we can distinguish these models using the precision measurement at the ILC.

Finally, we consider the small soft mass scenario.
In this scenario, for example, when $\tan\beta =3$, $\mu =180~\GeV$, $\mu_\Delta^{} =330~\GeV$, $\mu_S^{} =150~\GeV$, $\tilde{m}_\Delta^{} =100~\GeV$, $\tilde{m}_S^{} =300~\GeV$ and $m_{\rm SUSY}^{} =2~\TeV$, all masses of additional Higgs bosons are less than $500~\GeV$. 
Therefore we can directly produce and test the additional Higgs bosons at the ILC.
For instance, we can test the charged triplet $\Delta^\pm$ by the search of $e^+ e^- \rightarrow \Delta^+ \Delta^- \rightarrow tb \bar{t} \bar{b}$ process via $H^\pm -\Delta^\pm$ mixing.

\section{Conclusions}
We investigate phenomenology of the Higgs sector of the model of the supersymmetric grand Gauge-Higgs unification.
The SGGHU has adjoint chiral supermultiplets at the SUSY breaking scale.
In particular, the $SU(2)$ triplet and the singlet are included in the Higgs sector.
Since these adjoint chiral supermultiplets are unified the gauge field at the GUT scale, there is neither $\hat{S}\hat{S}\hat{S}$ nor $\hat{S}\hat{\Delta}\hat{\Delta}$ term, and trilinear couplings $\lambda_S^{}$ and $\lambda_\Delta^{}$ are related to the gauge coupling.
Therefore we can obtain the values of $\lambda_S$ and $\lambda_\Delta$ by solving the renormalization group equations.

The predicted values of the Higgs sector parameters can deviate from the MSSM and the SM by $O(1)~\% -O(10)~\%$. 
The strategy of discovering the SGGHU is the following. 
First, we test the mass of the charged Higgs boson and the heavy CP-even Higgs boson at the LHC. 
Then, we measure the coupling constants of the Higgs boson at the ILC. 
Through two steps, we can distinguish a model. 
The SGGHU is a good example of the GUT verifiable at colliders.

\begin{acknowledgments}
This paper is based on a work in collaboration with M. Kakizaki, S. Kanemura and T. Yamashita which is still in progress\cite{SGGHU}.
\end{acknowledgments}

\bigskip 

\end{document}